\documentclass[aps,prl,twocolumn,superscriptaddress,showpacs]{revtex4}
\usepackage{graphicx,graphics,amssymb,amsmath,hyperref}

\begin{document}

\def\etal#1{ {\em et al.}}
\def\tit#1{}
\def\jour#1{#1}

\title{Quantum-critical spin dynamics in quasi-one-dimensional antiferromagnets}

\author{S. Mukhopadhyay}
\affiliation{Laboratoire National des Champs Magn\'etiques Intenses, LNCMI - CNRS (UPR3228), UJF, UPS and INSA, BP 166, 38042 Grenoble Cedex 9, France}

\author{M. Klanj\v{s}ek}
\email{martin.klanjsek@ijs.si}
\affiliation{Laboratoire National des Champs Magn\'etiques Intenses, LNCMI - CNRS (UPR3228), UJF, UPS and INSA, BP 166, 38042 Grenoble Cedex 9, France}
\affiliation{Jo\v{z}ef Stefan Institute, Jamova 39, 1000 Ljubljana, Slovenia}
\affiliation{EN-FIST Centre of Excellence, Dunajska 156, 1000 Ljubljana, Slovenia}

\author{M. S. Grbi\'c}
\affiliation{Laboratoire National des Champs Magn\'etiques Intenses, LNCMI - CNRS (UPR3228), UJF, UPS and INSA, BP 166, 38042 Grenoble Cedex 9, France}
\affiliation{Department of Physics, Faculty of Science, University of Zagreb, PP 331, HR-10002 Zagreb, Croatia}

\author{R. Blinder}
\affiliation{Laboratoire National des Champs Magn\'etiques Intenses, LNCMI - CNRS (UPR3228), UJF, UPS and INSA, BP 166, 38042 Grenoble Cedex 9, France}

\author{H. Mayaffre}
\author{C. Berthier}
\affiliation{Laboratoire National des Champs Magn\'etiques Intenses, LNCMI - CNRS (UPR3228), UJF, UPS and INSA, BP 166, 38042 Grenoble Cedex 9, France}

\author{M. Horvati\'c}
\email{mladen.horvatic@lncmi.cnrs.fr}
\affiliation{Laboratoire National des Champs
Magn\'etiques Intenses, LNCMI - CNRS (UPR3228), UJF, UPS and INSA, BP 166, 38042 Grenoble Cedex 9, France}

\author{M. A. Continentino}
\affiliation{Centro Brasileiro de Pesquisas F\'isicas, 22290-180 Rio de Janeiro, Brazil}

\author{A. Paduan-Filho}
\affiliation{Universidade de S\~ao Paulo, 05315-970 S\~ao Paulo, Brazil}

\author{B. Chiari}
\author{O. Piovesana}
\affiliation{Dipartimento di Chimica, Universit\'a di Perugia, I-06100, Perugia, Italy}

\date{\today}



\begin{abstract}
By means of nuclear spin-lattice relaxation rate $T_1^{-1}$, we follow the spin dynamics as a function of the applied
magnetic field in two gapped one-dimensional quantum antiferromagnets: the anisotropic spin-chain system
NiCl$_2$-4SC(NH$_2$)$_2$ and the spin-ladder system (C$_5$H$_{12}$N)$_2$CuBr$_4$. In both systems, spin excitations are
confirmed to evolve from magnons in the gapped state to spinons in the gapples Tomonaga-Luttinger-liquid state. In
between, $T_1^{-1}$ exhibits a pronounced, continuous variation, which is shown to scale in accordance with quantum
criticality. We extract the critical exponent for $T_1^{-1}$, compare it to the theory, and show that this behavior is
identical in both studied systems, thus demonstrating the universality of quantum critical behavior.
\end{abstract}

\maketitle

\begin{figure}[b]
\includegraphics[width=1\linewidth]{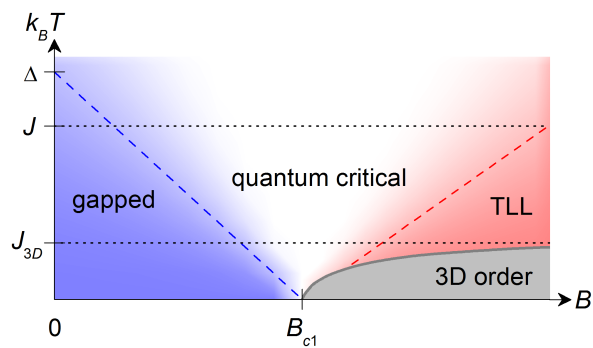}
\caption{(color online) Characteristic field-temperature phase diagram of weakly coupled, gapped antiferromagnetic
chains or ladders around the critical field $B_{c1}$ that closes the gap. The slopes $-g\mu_B$ and
$0.76g\mu_B$~\cite{Maeda_2007} of the temperature crossovers to the gapped and TLL regions, respectively, are indicated
by dashed lines. Dotted lines indicate characteristic temperature scales set by the dominant, 1D coupling $J$ and weak,
3D couplings $J_{3D}$.} \label{fig1}
\end{figure}

Quantum phase transitions are currently in the focus of condensed-matter
physics~\cite{Sachdev_1999,Coleman_2005,Sachdev_2011,Si_2010}. In contrast to classical phase transitions, driven by
thermal fluctuations, quantum phase transitions are driven by quantum fluctuations that can be tuned by non-thermal
control parameters, like the magnetic field, pressure or chemical composition. The influence of a quantum critical
point (QCP), where the continuous quantum phase transition occurs at zero temperature, extends to a broad V-shaped
region of quantum criticality at non-zero temperatures (like in Fig.~\ref{fig1}). A complex physics in this region is
universal, i.e., insensitive to the microscopic properties of the system, and scale invariant, with temperature setting
the only energy scale. Quantum phase transitions have been experimentally studied in heavy-fermion
metals~\cite{Si_2010,Schroder_2000}, magnetic insulators~\cite{Sebastian_2006,Ruegg1_2008,Thielemann1_2009,Coldea_2010}
and cold atoms~\cite{Hung_2011,Zhang_2012}. Magnetic insulators exhibit relatively simple and well-defined Hamiltonians
and allow for powerful local probes accessing spin statics and dynamics, like neutron
scattering~\cite{Ruegg1_2008,Thielemann1_2009,Coldea_2010} and nuclear magnetic resonance (NMR)~\cite{Klanjsek_2008}.
Nevertheless, a clear experimental demonstration of the quantum critical behavior in magnetic insulators is still
missing.

\begin{figure*}
\includegraphics[width=1\linewidth]{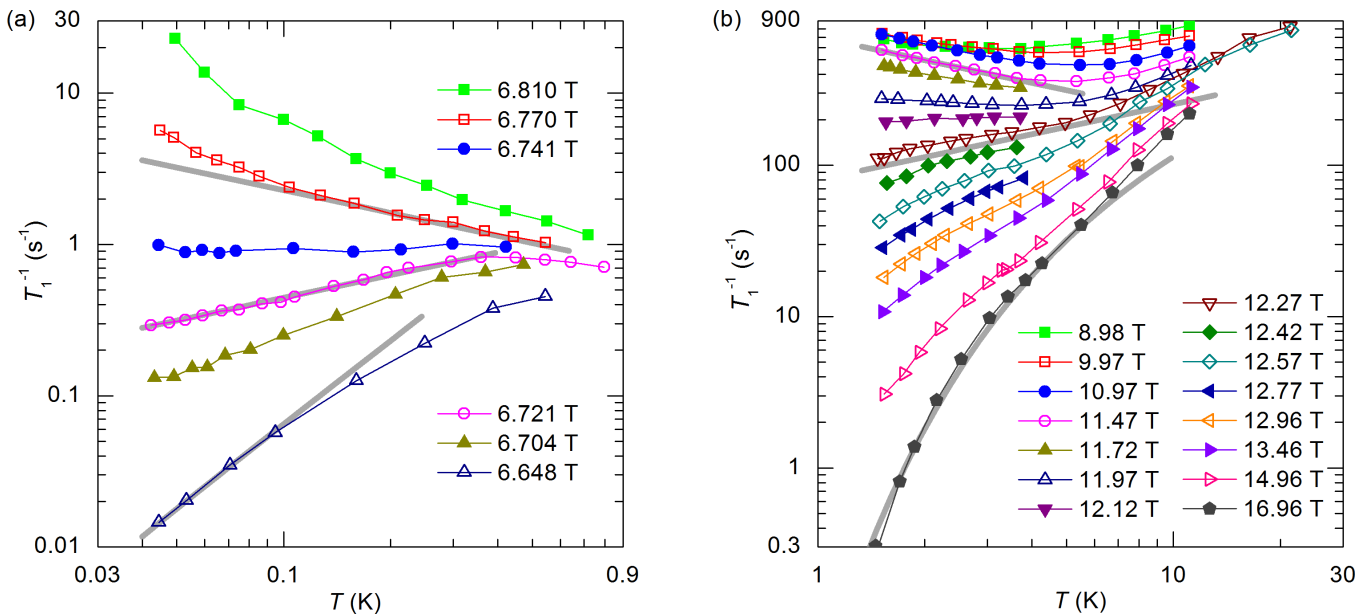}
\caption{(color online) $T_1^{-1}$ as a function of temperature $T$ for several magnetic field values around the QCP.
(a) In BPCB, $^{14}$N $T_1^{-1}$ data are taken around $B_{c1}=6.723$~T on N(1)$_{\rm II}$ NMR lines as defined in
Ref.~\cite{Klanjsek_2008}. The value of $B_{c1}$ is determined from the low-temperature magnetization as measured by
$^{14}$N NMR hyperfine shift. (A more precise determination provides a slightly higher value than reported in
Ref.~\cite{Klanjsek_2008}.) (b) In DTN, proton $T_1^{-1}$ data are taken around $B_{c2}=12.325$~T on the
highest-frequency NMR line. Magnetic field is aligned with the $c$ axis to within $1^\circ$. The value of $B_{c2}$ is
determined from the low-temperature boundary $T_c(B)$ of the 3D ordered state as measured by proton
NMR~\cite{Mukhopadhyay_2012}. Thick gray lines in (a) and (b) are $T_1^{-1}(T)$ predictions for the TLL behavior close
to the critical field ($T_1^{-1}\propto T^{-1/2}$), for the tentative quantum critical behavior exactly at the critical
field ($T_1^{-1}\propto T^{1/2}$), and for the gapped behavior, from top to bottom, respectively.} \label{fig2}
\end{figure*}

Particularly convenient are systems of weakly coupled one-dimensional (1D) gapped antiferromagnets~\cite{Sachdev_1994},
like spin chains or ladders, for two reasons. First, the applied magnetic field $B$ simply controls the gap between the
ground state and the lowest spin excitations, i.e., magnons. If $\Delta$ is the zero-field gap, this gap closes at the
critical field $B_{c1}=\Delta/(g\mu_B)$, which defines the QCP ($\mu_B$ is Bohr magneton). Beyond this QCP, magnons
fractionalize into pairs of spinons~\cite{Thielemann1_2009}, characteristic of the gapless, Tomonaga-Luttinger-liquid
(TLL) ground state~\cite{Klanjsek_2008,Giamarchi_2004}. Second, weakly coupled spin chains or ladders feature well
separated temperature scales relevant for the physics in 3D and 1D. These are characterized respectively by the
corresponding weak, 3D exchange couplings $J_{3D}$ and the dominant, 1D exchange coupling $J$ (Fig.~\ref{fig1}). In
particular, in the temperature range $k_BT<J_{3D}$ ($k_B$ is Boltzmann constant), 3D couplings lead to the 3D ordered
state in the TLL region~\cite{Klanjsek_2008}. In the range $J_{3D}<k_BT<J$, the physics gradually becomes 1D, while
above $J$ the 1D identity of the system is lost. As sketched in Fig.~\ref{fig1}, crossover temperatures to both the
gapped and TLL regions depend linearly on $B-B_{c1}$ around the QCP~\cite{Maeda_2007}. They outline a characteristic
V-shaped quantum critical region in the field-temperature phase diagram~\cite{Ruegg2_2008}. Spin dynamics in this
region has not been systematically explored yet.

We explore the quantum-critical spin dynamics in two particularly clean and convenient model systems:
NiCl$_2$-4SC(NH$_2$)$_2$ (DTN) containing chains of $S=1$ spins subject to a single-ion
anisotropy~\cite{Paduan_1981,Zapf_2006,Zvyagin_2007}, and (C$_5$H$_{12}$N)$_2$CuBr$_4$ (BPCB) containing spin-$1/2$
ladders~\cite{Thielemann1_2009,Klanjsek_2008,Ruegg2_2008,Watson_2001,Thielemann2_2009}. Both compounds feature
experimentally accessible critical fields, while the crystal symmetry assures the absence of the antisymmetric
Dzyaloshinskii-Moriya interaction that could perturb the closing of the gap at the QCP. In the strong-coupling
approximation~\cite{Klanjsek_2008}, where only the two spin states that close the gap are kept (among four rung states
for the ladder and three spin states for the chain), both systems are described by the same effective spin-$1/2$ $XXZ$
chain model with the coupling anisotropy $J_Z/J_{X,Y}=0.5$. Accordingly, they exhibit a similar phase diagram, which
contains also the second QCP at the critical field $B_{c2}$ that separates the gapless state from a gapped, fully
polarized state~\cite{Zapf_2006,Zvyagin_2007,Thielemann1_2009,Klanjsek_2008,Ruegg2_2008,Watson_2001,Thielemann2_2009}.
While the 1D couplings are comparable in both compounds, their 3D couplings differ by an order of magnitude
(Table~\ref{tab1}). A comparison of the quantum critical behavior in both compounds thus offers a severe test of
universality. Indeed, our results allow us to demonstrate (i) the universality and (ii) scale invariance of the
quantum-critical spin dynamics, and (iii) extract the critical exponent for $T_1^{-1}$, which is compared to the
existing theoretical predictions~\cite{Orignac_2007}.

\begin{table}[b]
    \begin{ruledtabular}
        \begin{tabular}{cccccc}
            ~ & $J_0$ & $J$ & $J/J_0$ & $J_{3D}$ & $T_{3D}^{\rm max}$  \\ \hline
            BPCB & 12.9 & 3.6 & 0.28 & 0.08 & 0.11 \\
            DTN & 8.9 & 4.4 & 0.49 & 0.72 & 1.2 \\
        \end{tabular}
    \end{ruledtabular}
\caption{The exchange couplings in the effective spin-$1/2$ $XXZ$ chain model for BPCB (DTN): on-site coupling $J_0$ is
the rung coupling $J_\perp$ (single-ion anisotropy $D$), 1D coupling $J$ is the leg coupling $J_\parallel$ (intrachain
coupling $2J_c$), 3D coupling $J_{3D}$ is the interladder coupling $zJ'$ with $z=4$ (interchain coupling $zJ_{a,b}$
with $z=4$), and the highest transition temperature $T_{3D}^{\rm max}$ to the 3D ordered state, all in kelvin
units~\cite{Klanjsek_2008,Zvyagin_2007}.} \label{tab1}
\end{table}

Nuclear spin-lattice relaxation rate $T_1^{-1}$ measurements allow us to monitor the spin dynamics around the QCP in
both compounds to high precision and in great detail. Namely, $T_1^{-1}$ can cover many orders of magnitude keeping the
same precision, while the magnetic field $B$ as a tuning parameter can be easily controlled. $T_1^{-1}$ provides a
direct access to the low-energy spin excitations, as it probes nearly zero-energy limit of the local (i.e., momentum
integrated) spin-spin correlation function~\cite{Moriya_1956}. Fig.~\ref{fig2} shows $T_1^{-1}(T)$ datasets for various
magnetic field values around the QCP. In BPCB, the datasets taken around $B_{c1}$ exhibit a power-law behavior,
$T_1^{-1}\propto T^\alpha$ (i.e., they are linear in a log-log scale), over nearly a decade [Fig.~\ref{fig2}(a)]. The
exponent $\alpha$ varies rapidly across the critical field, resulting in a fan-like pattern of the data. At low
temperature, the power-law (i.e., linear) behavior is modified at the lowest field value by the gap opening, which
reduces $T_1^{-1}$, and at the two highest field values by the 3D critical fluctuations, which enhance $T_1^{-1}$ close
to the boundary $T_c(B)$ of the 3D ordered state~\cite{Klanjsek_2008}. At high temperature, the deviation from the
power-law behavior starts above $0.3$~K, where all the datasets assume a decreasing trend. After passing a broad
minimum starting at $4$~K, which is comparable to $J$ in BPCB (Table~\ref{tab1}), the relaxation starts to increase
towards the high-temperature paramagnetic limit (not shown). Because of the low $B_{c1}$ value in DTN, the
$T_1^{-1}(T)$ data are instead taken around $B_{c2}$, and they exhibit essentially the same power-law fan-like pattern
as in BPCB [Fig.~\ref{fig2}(b)], although limited to a narrower temperature range below $4$~K. Above this temperature,
which is comparable to $J$ in DTN (Table~\ref{tab1}), the relaxation directly starts to increase to the
high-temperature paramagnetic regime. The low-temperature behavior again reflects entering into the gapped, this time
fully polarized region at the high field values, while the divergence of relaxation at the low field values, close to
$T_c(B)$, is not visible, as the lowest covered temperature of $1.5$~K is not close enough to the 3D ordered state
[$T_c(10\,{\rm T})=0.9\,{\rm K}$].

The limits of this fan-like pattern in Figs.~\ref{fig2}(a) and (b) are easily understood and reflect the nature of the
low-energy spin excitations on each side of the QCP. Namely, in the gapless, TLL region, the two-spinon continuum in a
dynamic correlation function leads to the power-law behavior $T_1^{-1}\propto T^{1/(2K)-1}$, where $K$ is the TLL
exponent~\cite{Klanjsek_2008,Giamarchi_1999}. Approaching the QCP from the TLL side, $K$ gradually increases to $1$,
meaning that $\alpha=1/(2K)-1$ decreases to $-1/2$, which is the lowest expected value of $\alpha$. As indicated by
upper gray lines in Figs.~\ref{fig2}(a) and (b), this lowest $\alpha$ value is indeed observed just above $B_{c1}$ in
BPCB, for $k_BT>J_{3D}$, and just below $B_{c2}$ in DTN. In the gapped region, magnon excitations over the gap result
in an activated behavior $T_1^{-1}\propto T^{\alpha_0}\exp[\pm g\mu_B(B-B_{c1,2})/(k_BT)]$, where $g=2.176$ for
BPCB~\cite{Klanjsek_2008} and $2.26$ for DTN~\cite{Zvyagin_2007}, the $+$ sign refers to $B_{c1}$ and $-$ to $B_{c2}$.
The exponent $\alpha_0$ depends on the effective dimension of the magnon dispersion relation as selected by thermal
fluctuations $k_BT$. On raising temperature, $\alpha_0$ gradually decreases from $2$ for $k_BT<J_{3D}$ (3D case)
towards $0$ for $J_{3D}\ll k_BT<J$ (1D case). The $T_1^{-1}(T)$ datasets deepest in the gapped region indeed exhibit
this type of low-temperature behavior, with $\alpha_0=1.8$ for BPCB (calculated at $0.05$~K) and $0.83$ for DTN
(calculated at $2$~K), as shown by lower gray lines in Figs.~\ref{fig2}(a) and (b). In between the gapped and TLL
regions neither description applies, implying that in the white region of Fig.~\ref{fig1} spin excitations are neither
spinons nor magnons. Interestingly, these excitations lead approximately to $T_1^{-1}\propto T^{1/2}$ exactly at the
critical field in both compounds, as indicated by middle gray lines in Figs.~\ref{fig2}(a) and (b).

\begin{figure}
\includegraphics[width=1\linewidth]{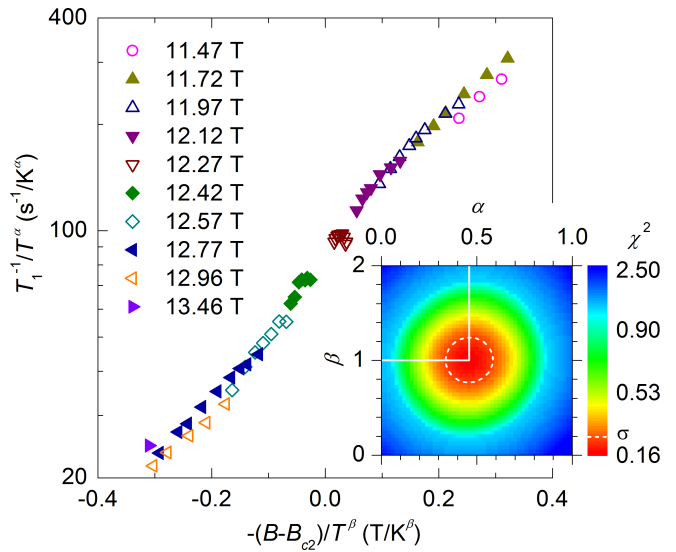}
\caption{(color online) Demonstration of the scale invariance in the quantum critical region of DTN. The best collapse
of different $T_1^{-1}(T)$ datasets [displayed in Fig.~\ref{fig2}(b)] on the same curve in scaling variables
$-(B-B_{c2})/T^\beta$ and $T_1^{-1}/T^\alpha$ is obtained for the critical exponents $\alpha=0.46\pm 0.12$ and
$\beta=1.00\pm 0.24$, with mean values used in the plot. Inset shows the color plot of $\chi^2(\alpha,\beta)$ measuring
the goodness of this collapse. Obtained mean values are indicated by solid white lines, their uncertainties are given
by a standard deviation $\sigma$ (dashed white contour line).} \label{fig3}
\end{figure}

To show that spin excitations in the white region of Fig.~\ref{fig1} are characteristic of quantum criticality, we
establish a scaling relation for the $T_1^{-1}$ data in this region. Assuming that $T_1^{-1}$ and the parameter
controlling the proximity to a QCP, $\pm(B-B_{c1,2})$, scale as powers $\alpha$ and $\beta$ of temperature, the scaling
relation reads
\begin{equation}\label{eq_T1}
    \frac{T_1^{-1}}{T^\alpha}=F\left[\frac{\pm(B-B_{c1,2})}{T^\beta}\right],
\end{equation}
where $F$ is the scaling function. To check Eq.~(\ref{eq_T1}), we first focus on DTN. We crop the ranges of the
$T_1^{-1}(T)$ datasets by the constraints $g\mu_B\vert B-B_{c2}\vert <0.5\,k_BT$ and $T<4$~K to confine them well to
the quantum critical region where the power-law behavior is observed. Then we look for the values of the exponents
$\alpha$ and $\beta$ leading to the best collapse of the cropped datasets on the same curve in scaling variables
$-(B-B_{c2})/T^\beta$ and $T_1^{-1}/T^\alpha$. For a given pair of $\alpha$ and $\beta$ values, we fit the dataset
containing \emph{all} the cropped and scaled datasets by the appropriate analytical function, a third-order polynomial
in our case, and use the $\chi^2$ of this fit as a measure of the collapse. As shown in Fig.~\ref{fig3} inset, the
minimization of $\chi^2$ as a function of both exponents leads to $\alpha=0.46\pm 0.12$ and $\beta=1.00\pm 0.24$. A
corresponding excellent collapse of all the datasets on the same curve, shown in Fig.~\ref{fig3}, provides a nice
demonstration of scale invariance. The energy scale is set only by temperature, and the obtained \emph{linear} scaling
of the control parameter $-(B-B_{c2})$ with temperature (i.e., $\beta=1$) is a clear sign of quantum
criticality~\cite{Sachdev_1999,Coleman_2005,Sachdev_2011}. The same analysis for BPCB, with $T_1^{-1}(T)$ datasets
confined to the range between $0.1$~K ($\sim J_{3D}/k_B$) and $0.3$~K, gives similar values $\alpha=0.48\pm 0.06$ and
$\beta=1.04\pm 0.08$. In both cases, Eq.~(\ref{eq_T1}) gives $T_1^{-1}\propto T^\alpha$ with the critical exponent
$\alpha\approx 1/2$ \emph{exactly} at the critical field. This result differs from the existing theoretical predictions
for the 3D quantum criticality, $\alpha=3/4$, and for the 1D quantum criticality, $\alpha=-1/2$, both given in
Ref.~\cite{Orignac_2007}. In the white region in Fig.~\ref{fig1}, the first case applies to the range $k_BT<J_{3D}$.
This region separates the gapped state from the 3D ordered state, which can be understood as a Bose-Einstein condensate
(BEC) of magnons~\cite{Giamarchi_1999,Nikuni_2000,Giamarchi_2008}. The second case applies to the range $J_{3D}\ll
k_BT<J$, where the effect of $J_{3D}$ is negligible. However, our result applies to the middle of the range
$J_{3D}<k_BT<J$, which has not been theoretically described yet.

\begin{figure}
\includegraphics[width=1\linewidth]{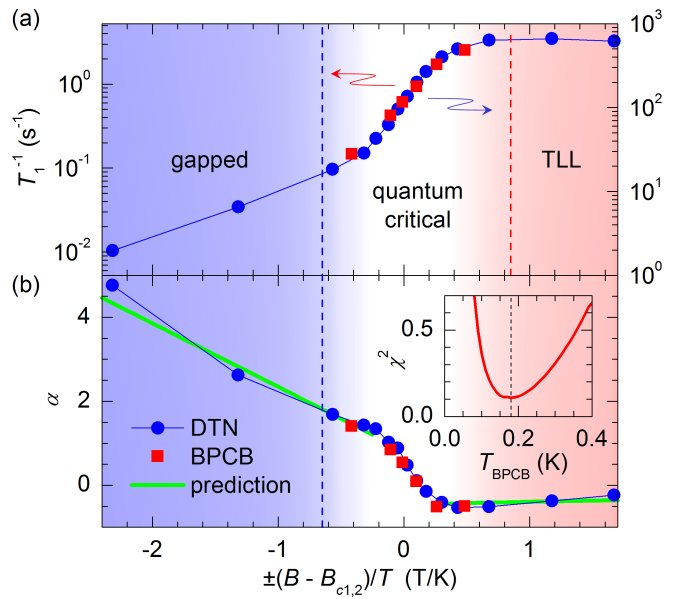}
\caption{(color online) Demonstration of the universality of quantum-critical behavior. (a) $T_1^{-1}$ versus
$\pm(B-B_{c1,2})/T$ datasets for BPCB at $0.18$~K and DTN at $2$~K [determined from the data in Figs.~\ref{fig2}(a) and
(b)] overlap perfectly in the quantum critical region. (b) Corresponding overlap of the $\alpha$ datasets, where the
power-law exponents $\alpha$ are evaluated as the slopes of tangents to the $T_1^{-1}(T)$ datasets in a log-log scale.
Inset shows the $\chi^2(T_{\rm BPCB})$ plot measuring the goodness of this overlap at a fixed $T_{\rm DTN}=2$~K,
defining the optimal $T_{\rm BPCB}=0.18$~K (indicated by the dashed line). Dashed lines in (a) and (b) indicate the
crossover temperatures to the gapped (blue) and TLL (red) regions, as in Fig.~\ref{fig1}, and the green lines in (b)
are $\alpha$ versus $\pm(B-B_{c1,2})/T$ predictions in these two regions. Different $T_1^{-1}$ scales in (a) are due to
different gyromagnetic ratios and different NMR hyperfine couplings for $^{14}$N and protons.} \label{fig4}
\end{figure}

Finally, we show that the similarity of the $T_1^{-1}$ fan-like patterns in both compounds is not only qualitative but
also quantitative. For this purpose, from the data displayed in Fig.~\ref{fig2} we extract the \emph{field} variations
of $T_1^{-1}$ and of the (effective) power-law exponent $\alpha=\partial\ln(T_1^{-1})/\partial\ln(T)$ at a given
temperature. Having established $\pm(B-B_{c1,2})/T$ as the proper scaling variable in the quantum critical region, we
plot in Fig.~\ref{fig4} $T_1^{-1}$ and $\alpha$ in DTN as a function of this variable at $T_{\rm DTN}=2$~K. Strong
variations of both observables are localized in a narrow range around the QCP, where $T_1^{-1}$ increases by a factor
of $10$ and $\alpha$ changes from $1.5$ to $-0.5$ from the gapped to the TLL region. Outside the quantum critical
region, the observed $\alpha$ versus $\pm(B-B_{c1,2})/T$ variation is nicely reproduced [see Fig.~\ref{fig4}(b)] on the
basis of $T_1^{-1}$ expressions given above: with $\alpha=\pm g\mu_B(B-B_{c1,2})/(k_BT)+\alpha_0$ on the gapped side,
where $\alpha_0=0.83$ (calculated at $T_{\rm DTN}$), and with $\alpha=1/(2K)-1$ on the TLL side, where we use $K(B)$ in
the strong-coupling approximation~\cite{Klanjsek_2008}. In the quantum critical region, where neither description
applies, we look for the temperature $T_{\rm BPCB}$ to achieve the best overlap of the $T_1^{-1}$ and $\alpha$ datasets
for BPCB with those for DTN. As shown in Fig.~\ref{fig4}(b) inset, the obtained $T_{\rm BPCB}=0.18$~K leads to $T_{\rm
DTN}/T_{\rm BPCB}=11$. We get the same ratio for any chosen $T_{\rm DTN}$ in the covered temperature range. An
excellent overlap of the BPCB datasets with those for DTN, despite different exchange couplings defining their
Hamiltonians (Table~\ref{tab1}), provides a clear demonstration of universality. Spin dynamics in the quantum critical
region can be understood in terms of strongly interacting magnons~\cite{Orignac_2007}. The characteristic temperature
for the magnon-magnon interaction is given by the highest transition temperature $T_{3D}^{\rm max}$ to the 3D ordered
(i.e., BEC) state (see Fig.~\ref{fig1} and Table~\ref{tab1}). The fact that $T_{3D,{\rm DTN}}^{\rm max}/T_{3D,{\rm
BPCB}}^{\rm max}=11$ precisely corresponds to the obtained $T_{\rm DTN}/T_{\rm BPCB}$ indicates that the universality
is defined by an interaction-dependent scale factor~\cite{Sachdev_1994,Stauffer_1972}. In the end, we note that a
dataset like those plotted in Fig.~\ref{fig4}(a) was obtained for the spin-chain compound CuPzN in
Ref.~\cite{Kuhne_2011}, but was interpreted within the TLL framework, which does not apply to the quantum critical
region.

In summary, we showed that the quantum-critical spin dynamics in gapped quasi-1D antiferromagnets cannot be understood
in terms of spinons or magnons, but rather in terms of more complicated spin excitations. Their behavior was
experimentally demonstrated to be scale invariant and universal, where the scale factor is defined by the magnon-magnon
interaction. We extracted the critical exponent for $T_1^{-1}$ in the region which is not covered by any theory. For
the well developed 3D region at lower temperatures ($k_BT\ll J_{3D}$) and for the 1D region at higher temperatures
($J_{3D}\ll k_BT<J$) theoretical descriptions exist~\cite{Sachdev_1994,Orignac_2007}. As $T_1^{-1}$ can be expressed in
terms of the dynamical susceptibility~\cite{Moriya_1956}, the extension of our experimental study to these regions
should allow to extract the universal critical exponents for susceptibility and correlation length in 1D and 3D, and
compare them to existing theories.

We acknowledge fruitful discussions with T. Giamarchi. Part of this work has been supported by the French ANR project
NEMSICOM, by EuroMagNET network under the EU contract No. 228043, by the ARRS project No. J1-2118, and by the EU FP7
project SOLeNeMaR N° 229390.

\end{document}